# Gating a single-molecule transistor with individual atoms

Jesús Martínez-Blanco[1], Christophe Nacci[1], Steven C. Erwin[2], Kiyoshi Kanisawa[3], Elina Locane[4], Mark Thomas[4], Felix von Oppen[4], Piet W. Brouwer[4] and Stefan Fölsch[1]

[1] Paul-Drude-Institut für Festkörperelektronik, Hausvogteiplatz 5-7, 10117 Berlin, Germany.
[2] Center for Computational Materials Science, Naval Research Laboratory, Washington, DC 20375, USA.
[3] NTT Basic Research Laboratories, NTT Corporation, Atsugi, Kanagawa, 243-0198, Japan.
[4] Dahlem Center for Complex Quantum Systems and Fachbereich Physik, Freie Universität Berlin, 14195 Berlin, Germany.

**Transistors, regardless of their size, rely on electrical gates to control the conductance between source and drain contacts. In atomic-scale transistors, this conductance is exquisitely sensitive to single electrons hopping via individual orbitals[1,2]. Single-electron transport in molecular transistors has been previously studied using top-down approaches to gating, such as lithography and break junctions[1,3,4,5,6,7,8,9,10,11]. But atomically precise control of the gate – which is crucial to transistor action at the smallest size scales – is not possible with these approaches. Here, we used individual charged atoms, manipulated by a scanning tunnelling microscope,[12] to create the electrical gates for a single-molecule transistor. This degree of control allowed us to tune the molecule into the regime of sequential single-electron tunnelling, albeit with a conductance gap more than one order of magnitude larger than observed previously[8,11,13,14]. This unexpected behaviour arises from the existence of two different orientational conformations of the molecule, depending on its charge state. Our results show that strong coupling between these charge and conformational degrees of freedom leads to new behaviour beyond the established picture of single-electron transport in atomic-scale transistors.**

We created molecular transistors using either phthalocyanine ($H_2Pc$) or copper phthalocyanine (CuPc) molecules, depicted in Fig. 1a. These are planar π-conjugated molecules which readily adsorb on an InAs(111)A substrate[15] at precisely defined locations above the In-vacancy site of the $2 \times 2$ reconstructed surface (Fig. 1b). The molecule is physisorbed on the surface by van der Waals interaction with no significant charge transfer. Due to weakness of this binding, the adsorbed molecules can be readily repositioned among different vacancy sites by using the tip of a cryogenic scanning tunnelling microscope (STM).

The InAs(111)A substrate surface is characterized by a low concentration of native defects, +1 ionized indium adatoms ($In_{ad}$). The adatoms render the surface weakly metallic with the Fermi level pinned in the conduction band[16]. They can also be repositioned by the STM tip[17]. By arranging a group of charged adatoms in a carefully chosen configuration, we designed and created a variety of electrostatic surface-potential landscapes for the nearby region with sub-angstrom accuracy.

This positional control over both the charged indium adatoms and the phthalocyanine molecules allows us, first, to design and create an atomically precise potential landscape using the $In^{+1}$ adatoms and, then, to position the molecule within that landscape. The potential landscape provides the transistor gate[18,19] by acting on the electron energy levels of the weakly coupled molecule. To complete the three-terminal transistor structure, the STM tip and InAs substrate serve as source and drain contacts, respectively. Electrostatic gating by repositionable charged defects has previously been used to shift the binding energy of dopants near a GaAs surface[20]. Here, we apply this concept to control the charge state of a single molecule (see supplementary information). Figure 1c,d shows topographic images of two $In_{ad}$ trimers with an $H_2Pc$ molecule in a remote position (c), and midway between the trimers (d). The molecule is imaged as a uniform protrusion because it performs rapid rotational jumps at sample biases $V_b$ larger than $\pm 0.1$ V[15]. In the midpoint position (d), the electrostatic gating is sufficient to shift the lowest-unoccupied molecular orbital (LUMO) below the Fermi level $E_F$, charging the molecule negatively. As a consequence, the local tunnelling barrier[21,22] is increased, which reduces the apparent molecular height compared to the neutral state. To quantify the charge state we measured the electrostatic potential at the tip position[23,24] along the dashed line in Fig. 1c (Supplementary Figure 1). Figure 1e shows the difference in potentials obtained in the presence and absence of the molecule. Clearly, the potential difference is essentially zero when the molecule is located in the remote position (green dots), confirming its neutral state there. At the midpoint position (red squares), on the other hand, the charge state is −1, as confirmed by comparison with the calculated potential (blue curve) of a point charge screened by the surface-accumulated electrons[23,24,25] (see supplementary information).

To investigate the gating process systematically, we constructed $In_{ad}$ corrals that allowed us to fine-tune the gating through a linear potential landscape (Supplementary Fig. 2). We found that both CuPc and $H_2$Pc can be tuned into a regime of charge bistability. STM imaging in the bistable regime revealed that the two molecular charge states have different orientational conformations in their respective ground states. The left STM image in Fig. 2b shows a neutral $H_2$Pc in one of its three equivalent conformers, $A$, in which the molecular lobes are oriented parallel/perpendicular to one of the three $\langle 110 \rangle$ in-plane directions[15]. Upon charging, the molecule performs an in-plane rotation either clockwise or counter-clockwise (centre and right STM images, respectively). There are three equivalent conformers, $A$, in the neutral state and six equivalent conformers, $B$, in the negative charge state. This suggests that the transition $A \rightarrow B$ consists of an in-plane rotation by $\pm 15°$, as well as a small lateral displacement and/or tilt[26] of the molecule (Supplementary Fig. 3). Topographic images analogous to those in Fig. 2b were obtained for CuPc in its neutral and negatively charged state, respectively. The bistability observed here involves a coupled switching in charge and orientational conformation. We note that STM-driven conformational switching in adsorbed phthalocyanine molecules has also been reported by others.[27]

The electrostatic gating of the LUMO level strongly modifies the tunnelling conductance, as revealed by the current-*versus*-bias ($I$-$V_b$) curves in Fig. 3a, which were recorded over an extended bias range with the STM tip held fixed above a single CuPc molecule. The conductance gap is reduced as the level is brought closer to $E_F$, consistent with the simple picture of transport via a single unoccupied level within a double-barrier tunnel junction (DBTJ)[28,29]. This picture is also supported by the normalized differential conductance map in Fig. 3b, which was extracted from $I$-$V_b$ curves consecutively recorded at different values of the gate potential $\phi$. At a given value, the differential conductance is sharply peaked at biases $V_b$ for which the LUMO enters the energy window between the tip and sample Fermi levels, which leads to sequential tunnelling through the DBTJ. However, near the charge degeneracy point there is a conductance gap, ~600 mV, which is much larger than for conventional sequential tunnelling through a gated quantum dot. This gap is substantially greater than normally found, for example, in electron transport through single molecules in the Franck-Condon blockade regime of strong electron-vibron coupling[8,11,13,14].

The general features of the measured tunnelling conductance can be understood within a Coulomb blockade model[30] for the DBTJ, by treating the molecule as a quantum dot attached to the STM tip and the InAs(111)A surface, which act as electron reservoirs. Motivated by the experimental observations, we assume that the molecule can exist in two conformations, $A$ and $B$, and that each conformation can be either neutral or charged. Conformation $A$ is the ground state of the neutral molecule and $B$ the ground state of the charged molecule, as shown in Fig. 3c. We find that this assumption naturally explains the occurrence of the conductance gap in the $V_b$-$\phi$ plane. Indeed, both conformations give rise to a separate charge degeneracy point. The charge degeneracy point for $A$ is shifted to higher gate voltages relative to that for $B$. This defines several regions in the $V_b$-$\phi$ plane, labelled I-IX in Fig. 3d. In regions I-III the molecule is blockaded with respect to both conformations and hence conduction is suppressed, while in regions IV and V the molecule is not blockaded and so conduction occurs. The remaining regions, VI-IX, are more subtle as they are conducting with respect to one conformation but blockaded with respect to the other. Franck-Condon physics strongly suppresses tunnelling-induced switching between the conformations and therefore $A$ is stable in regions VI and VII, while $B$ is stable in VIII and IX. These correspond to the blockaded conformations and hence there is a conductance gap for all potentials $\phi$, consistent with the experimental observations.

The bistability in charge and conformation shown in Fig. 2 occurs within the conductance gap region where sequential tunnelling is blocked. Figure 4a shows $I$-$V_b$ curves recorded close to $E_F$ with the STM tip held fixed above an $H_2Pc$ molecule at the indicated $\phi$ values. Hysteresis in $I$-$V_b$ is easily seen at $\phi = 121$ mV (red), revealing bistable switching between two $I$-$V_b$ curves of different slope around $E_F$. The smaller $I$-$V_b$ slope is associated with the negative charge state, concomitant with the increase in local tunnelling barrier upon charging. Similar hysteretic behaviour was previously reported for the bistable charge switching of phthalocyanine molecules on NaCl layers grown on Cu substrates of suitable workfunction[21]. A slight detuning of the potential quenches the $I$-$V_b$ hysteresis in Fig. 4a, as shown by the blue $I$-$V_b$ curves, reflecting a dynamical crossover between the two states as detailed in the inset. Figure 4b shows the location of the hysteretic switching (red) and the crossover (blue) in the $V_b$-$\phi$ plane, together with the onset of sequential tunnelling via the LUMO (green). We observed similar behaviour for CuPc (Supplementary Fig. 4), for which the bistability occurs at a gating potential ~50 mV larger than for $H_2Pc$. Finally, the hysteretic switching dynamics is depicted in Fig. 4c showing that the coercivity (defined as the half width of the hysteresis) sharply rises and eventually saturates as the bias ramping speed is increased.

Based on the model description outlined in Fig. 3c,d, we set up a Master equation for the coupled electronic and conformational dynamics (see supplementary information for details). As shown in Fig. 4d, the numerical solution indeed predicts a pronounced conductance gap for all $\phi$, in agreement with the measurements in Figs. 3b and 4b. This model also explains the experimentally observed hysteretic behaviour in Fig. 4a and Supplementary Fig. 4. Region I in Fig. 4d consists of two sections, separated by the orange line, with different molecular ground states. Experimentally, these regions have different co-tunnelling conductances[30], which explains why we observe bistable $I$-$V_b$ characteristics (Fig. 4a,b). Moreover, as conformational changes are slow, the current can be hysteretic when varying parameters across this line. Experimentally, this is most pronounced at small biases (Fig. 4b), where electronic tunnelling (and hence the attempt rate for conformational switching) is weak. The observed coercivity as a function of ramping speed of the bias voltage (Fig. 4c) is consistent with the theoretical results from the Master equation as shown in Fig. 4e.

While our model qualitatively explains many features of the observations, some aspects may require a more refined description. For instance, the model predicts that the lines of current onset extrapolate to cross at the pertinent charge degeneracy point in the $V_b$-$\phi$ plane and thus at zero bias voltage. This is not actually the case in the experimental data in Figs. 3b and 4b. This effect might be due to finite temperature or to interactions beyond our constant-capacitance model (see supplementary information for further discussion).

We have shown that a single-molecule transistor can be electrostatically gated by arranging individual charged atoms with sub-angstrom precision using a scanning tunnelling microscope. The resulting transistor action reveals a conductance gap far larger than previously observed for Franck-Condon-blockaded transport[8,11,13,14], arising from strong coupling between charge and conformational degrees of freedom. Understanding and controlling this type of coupling – and the new kinds of behaviour to which it can lead – will be crucial for integrating atomic-scale transistors and other devices with existing semiconductor technologies.

## Methods Summary

**Experiment.** The measurements were carried out with an ultrahigh-vacuum (UHV) STM operated at a base temperature of 5 K. Electrochemically etched tungsten tips were cleaned in UHV by Ne ion sputtering and electron beam heating. Undoped InAs layers with thickness 20 nm were grown by molecular beam epitaxy (MBE) on an InAs(111)A wafer (purchased from *Wafer Technology Ltd*) to prepare the In-terminated InAs(111)A-(2×2) surface. After the MBE growth, the surface was capped with an amorphous layer of arsenic and transferred under ambient conditions to the UHV system of the STM apparatus. The As capping layer was removed by annealing at 630 K and the sample loaded into the microscope cooled down to 5 K. InAs(111)A samples prepared in this way showed the same surface features as MBE-grown and *in situ* investigated samples. $H_2Pc$ and CuPc specimens (purchased from *Aldrich*) were purified in UHV by repeated cycles of degassing. Low coverages (~1 × $10^{12}$ molecules $cm^{-2}$) were deposited directly into the microscope at sample temperatures <20 K by sublimation from a tantalum crucible held at 628 K. STM images were recorded in constant-current mode; bias voltages refer to the sample with respect to the tip. Spectroscopy measurements of the differential tunnelling conductance were performed by lock-in technique (10 mV peak-to-peak modulation at a frequency of 675 Hz) with disabled feedback loop.

**Theoretical calculations.** The theoretical model included four states of the phthalocyanine molecule, corresponding to the two conformations *A* and *B* and the two charge states for each conformation. The state of the molecule was described in terms of probabilities for each of these four states. In the weak tunnelling limit (tunnelling rate much smaller than temperature and spacing between the energies of the four molecular states) these probabilities can be determined from the solution of a Master equation that involves transition rates for thermally-induced transitions between different conformations of the molecule at constant charge, and transition rates between different charge states at the same conformation, due to tunnelling to/from the scanning tip and the substrate. The stationary solution of the Master equation was used to determine the onset of current flow in Fig. 4d; a time-dependent solution with a bias voltage that varies linearly in time was used to calculate the coercivity in Fig. 4e.

**References**

[1] Park, H., Park, J., Lim, A. K. L., Anderson, E. H., Alivisatos, A. P. & McEuen, P. L. Nanomechanical oscillations in a single-$C_{60}$ transistor. *Nature* **407**, 57-60 (2000).

[2] Kouwenhoven, L. P., Austing, D. G. & Tarucha, S. Few-electron quantum dots. *Rep. Prog. Phys.* **64**,701-736(2001).

[3] Liang, W., Shores, M. P., Bockrath, M., Long, J. R. & Park, H. Kondo resonance in a single-molecule transistor. *Nature* **417**, 725-729 (2002).

4 Kubatkin, S., Danilov, A., Hjort, M., Cornil, J., Brédas, J.-L., Stuhr-Hansen, N., Hedegård, P. & Bjørnholm, T. Single-electron transistor of a single organic molecule with access to several redox states. *Nature* **425**, 698-701 (2002).

5 Yu, L. H., Keane, Z. K., Ciszek, J. W., Cheng, L., Stewart, M. P., Tour, J. M. & Natelson, D. Inelastic electron tunneling via molecular vibrations in single-molecule transistors. Phys. Rev. Lett. **93**, 266802 (2004).

6 Roch, N., Florens, S., Bouchiat, V., Wernsdorfer W. & Balestro, F. Quantum phase transition in a single-molecule quantum dot. *Nature* **453**, 633-637 (2008).

7 Song, H., Kim, Y., Jang, Y. H., Jeong, H., Reed M. A. & Lee, T. Observation of molecular orbital gating. *Nature* **462**, 1039-1042 (2009).

8 Leturcq, R., Stampfer, C., Inderbitzin, K., Durrer, L., Hierold, C., Mariani, E., Schultz, M. G., von Oppen, F. & Ensslin, K. Franck–Condon blockade in suspended carbon nanotube quantum dots. *Nat. Phys*. **5**, 327-331 (2009).

9 Champagne, A. R., Pasupathy, A. N. & Ralph, D. C. Mechanically adjustable and electrically gated single-molecule transistors. *Nano Lett*. **5**, 305-308 (2005).

10 Perrin, M. L., Verzijl, C. J. O., Martin, C. A., Shaikh, A. J., Eelkema, R., van Esch, J. H., van Ruitenbeek, J. M., Thijssen, M. J., van der Zant, H. S. J. & Dulić, D. Large tunable image-charge effects in single-molecule junctions. *Nat. Nanotechnol*. **8**, 282-287 (2013).

11 Burzurí, E., Yamamoto, Y.,Warnock, M., Zhong, X., Park, K., Cornia, A. & van der Zant, H. S. J. Franck−Condon blockade in a single-molecule transistor. *Nano Lett*. **14**, 3191-3196 (2014).

12 Stroscio, J. A. & Eigler, D. M. Atomic and molecular manipulation with the scanning tunneling microscope. *Science* **254**, 1319-1326 (1991).

13 Koch, J. & von Oppen, F. Franck-Condon blockade and giant Fano factors in transport through single molecules. *Phys. Rev. Lett*. **94**, 206804 (2005).

[14] Ryndyk, D. A., Amico, P. D., Cuniberti, G. & Richter, K. Charge-memory effect in molecular junctions. Phys. Rev. B **78**, 085409 (2008).

[15] Nacci, C., Erwin, S.C., Kanisawa, K. & Fölsch S. Controlled switching within an organic molecule deliberately pinned to a semiconductor surface. *ACS Nano* **6**, 4190-4195 (2012).

[16] Olsson, L. Ö., Andersson, C. B. M., Håkansson, M. C., Kanski, J., Ilver, L. & Karlsson, U. O. Charge accumulation at InAs surfaces. *Phys. Rev. Lett.* **76**, 3626-3629 (1996).

[17] Fölsch, S., Yang, J., Nacci, C. & Kanisawa, K. Atom-by-atom quantum state control in adatom chains on a semiconductor. *Phys. Rev. Lett.* **103**, 096104 (2009).

[18] Piva, P. G., DiLabio, G. A., Pitters, J. L., Zikovsky, J., Rezeq, M., Dogel, S., Hofer, W. A. & Wolkow, R. A. Field regulation of single-molecule conductivity by a charged surface atom. *Nature* **435**, 658-661 (2005).

[19] Riss, A., Wickenburg, S., Tan, L. Z., Tsai, H.-Z., Kim, Y., Lu, J., Bradley, A. J., Ugeda, M. M., Meaker, K. L., Watanabe, K., Taniguchi, T., Zettl, A., Fischer, F. R., Louie, S. G. & Crommie, M. F. Imaging and tuning molecular levels at the surface of a gated graphene device. *ACS Nano* **8**, 5395-5401 (2014).

[20] Lee, D. H. & Gupta, J.A. Tunable field control over the binding energy of single dopants by a charged vacancy in GaAs. *Science* **330**, 1807-1810 (2010).

[21] Swart, I., Sonnleitner, T. & Repp, J. Charge state control of molecules reveals modification of the tunneling barrier with intramolecular contrast. *Nano Lett.* **11**, 1580-1584 (2011).

[22] Fernández-Torrente, I., Kreikemeyer-Lorenzo, D., Stróżecka, A., Franke, K.J. & Pascual, J.I. Gating the charge state of single molecules by local electric fields. *Phys. Rev. Lett.* **108**, 036801 (2012).

[23]Teichmann, K., Wenderoth, M., Loth, S., Ulbrich, R. G., Garleff, J.K., Wijnheijmer, A. P. & Koenraad P. M. Controlled charge switching on a single donor with a scanning tunneling microscope. *Phys. Rev. Lett.* **101**, 076103 (2008).

[24]Yang, J., Erwin, S. C., Kanisawa, K., Nacci, C. & Fölsch, S. Emergent multistability in assembled nanostructures. *Nano Lett.* **11**, 2486-2489 (2011).

[25]Stern, F. & Howard, W. E. Properties of semiconductor surface inversion layers in the electric quantum limit. *Phys. Rev.* **163**, 816-835 (1967).

[26]Schuler, B., Liu, W., Tkatchenko, A., Moll, N., Meyer, G., Mistry, A., Fox, D. & Gross, L. Adsorption geometry determination of single molecules by atomic force microscopy. *Phys. Rev. Lett.* **111**, 106103 (2013).

[27]Liljeroth, P., Repp, J. & Meyer, G. Current-induced hydrogen tautomerization and conductance switching of naphthalocyanine molecules. *Science* **317**, 1203-1206 (2007).

[28]Paulsson, M., Zahid, F. & Datta, S. Resistance of a molecule. *Nanoscience, engineering, and technology handbook*, edited by Goddard, W., Brenner, D., Lyshevski, S. & Iafrate, G. (CRC Press, Boca Raton, 2003).

[29]Nazin, G. V., Wu, S. W. & Ho, W. Tunneling rates in electron transport through double-barrier molecular junctions in a scanning tunneling microscope. *Proc. Nat. Acc. Sci.* **102**, 8832-8837 (2005).

[30]Nazarov, Y. V. & Blanter, Y. M. *Quantum transport* (Cambridge University Press, Cambridge 2009).

[31]Fölsch, S., Martínez-Blanco, J., Yang, J., Kanisawa, K. & Erwin, S.C. Quantum dots with single-atom precision. *Nat. Nanotechnol.* **9**, 505-508 (2014).

**Acknowledgements** This work was supported by the German Research Foundation (Collaborative Research Network SFB 658) and the Office of Naval Research through the Naval Research Laboratory's Basic Research Program. Some computations were performed at the DoD Major Shared Resource Center at AFRL.

Supplementary Information is linked to the online version of the paper at www.nature.com/nature.

**Contributions** J.M.-B., C.N., and S.F. performed the STM experiment and the experimental data analysis. K.K. performed the MBE growth of the InAs samples. S.C.E. predicted the charge-induced molecular reorientation based on density-functional theory calculations. E.L. and M.T. performed the generic model calculations of the coupled electronic and conformational dynamics. F.v.O. and P.W.B. developed the generic model. J.M.-B., S.C.E., E.L., M.T., F.v.O., P.W.B., and S.F. co-wrote the manuscript.

**Competing financial interests** The authors declare no competing financial interests.

**Corresponding authors** Correspondence and requests for materials should be addressed to S.F. (foelsch@pdi-berlin.de).

**Figure legends**

Fig. 1 **Electrostatic gating of an organic molecule using charged indium adatoms. a**, Structure of phthalocyanine ($H_2Pc$) and copper phthalocyanine (CuPc). **b**, Schematic model of InAs(111)A-(2×2) with In surface atoms in the topmost layer (green) and As atoms in the second layer (orange). **c**, STM topographic image (50 pA, 0.5 V) of an $H_2Pc$ molecule (left) and six In adatoms ($In_{ad}$), each charged +1, arranged as two trimers (right); the molecule is neutral and located $12a'$ ($a' = 8.57$ Å is the vacancy spacing) away from the midpoint between the trimers (origin of the $x$ axis, dashed line). **d**, Same as (c) after moving the molecule with the STM tip to the midpoint. The electrostatic potential of the $In_{ad}$ trimers charges the molecule negatively and reduces its apparent height from 2.2 Å (neutral) to 0.9 Å (negative). **e**, Difference in electrostatic potential $\Delta\phi(x)$ with and without the molecule measured at a tip height on the order of 6 Å along the dashed line in (c). For $H_2Pc$ in the remote position (green) $\Delta\phi(x)$ is essentially zero (consistent with $H_2Pc$ being neutral), whereas for $H_2Pc$ at the midpoint (red) $\Delta\phi(x)$ agrees well with the theoretical potential (blue curve) of a $-1$ point charge at the location of the molecule. The procedure to extract the experimental quantity $\Delta\phi$ is described in Supplementary Figure 1 by the example of the data point marked by the red arrow.

Fig. 2 **Change of molecular conformation upon charging. a**, STM images of $H_2Pc$ in the neutral (50 pA, 60 mV; left) and $-1$ charged state (50 pA, $-60$ mV; centre and right) recorded at intermediate gating of $\phi = 121$ mV. In the neutral state, the molecular lobes are oriented parallel/perpendicular to the $[10\overline{1}]$ in-plane direction. Upon charging, the molecule performs a

±15° in-plane rotation and a minor shift and/or tilt breaking the mirror symmetry relative to the

($\overline{1}$ 10) and (01 $\overline{1}$ ) plane, respectively. Coloured bars indicate the lobe orientations. Dark stripes in the left image are due to transient charging during the scanning. **b**, Schematic model of the molecular orientations corresponding to the cases in (a); for clarity, only the in-plane rotation is depicted.

Fig. 3 **Gap formation in the sequential tunnelling regime due to coupled charge and**
**conformational states. a**, Current-*versus*-bias ($I$-$V_b$) curves recorded over an extended bias range

with the STM tip held fixed above a single CuPc at the indicated gating potentials $\phi$; the

conductance gap changes as $\phi$ is varied. **b**, Normalized differential tunnelling conductance map

as a function of gating potential $\phi$ and sample bias $V_b$. The map corresponds to the stability

diagram of a gated quantum dot, but with a much larger residual gap in the region of charge

degeneracy. **c**, Allowed transitions among four different states (two conformations $A$ and $B$, in

two charge states 0 and −1) at low (upper panel), high (lower panel), and intermediate (middle)

gating potentials. **d**, Schematic stability diagram within a Coulomb blockade model for single-electron tunnelling. Red and blue lines are the Coulomb branches for the conformations $A$ and $B$, respectively. Transport through the molecule is blocked in regions I-III and regions VI-IX (white areas) and conducting in regions IV and V (grey areas).

Fig. 4 **Molecular charge bistability and switching dynamics within the conductance gap. a**,
$I$-$V_b$ curves measured with the STM tip fixed above a single $H_2Pc$ gated by surrounding $In_{ad}$

atoms; the potential $\phi$ at the molecular position is indicated. At intermediate $\phi$ (red) the molecular charge becomes bistable, manifested as $I$-$V_b$ hysteresis. Detuning in $\phi$ (blue) quenches the hysteresis and yields a dynamical crossover between two current levels (circles and inset). Arrows indicate the bias ramping direction. The dashed line marks an InAs surface state[31] confined by the ionized $In_{ad}$. **b**, Stability diagram of $H_2Pc$ showing that the crossover (blue) and hysteretic switching (red) observed in (a) occur within the conductance gap of sequential tunnelling via the LUMO. **c**, Coercivity of the hysteretic switching versus bias ramping speed for $H_2Pc$ (red) and CuPc (blue) measured at fixed gating potentials $\phi = 125$ mV and $\phi = 170$ mV, respectively. Error bars indicate the statistical variation extracted from $I$-$V_b$ curves consecutively recorded at fixed ramping speed. **d**, Calculated map of the tunnelling current $I(V_b, \phi)$. The orange line separates the $V_b$-$\phi$ plane into upper and lower regions, in which the molecule is predominantly in conformation $A$ and $B$, respectively. Dips in the co-tunnelling current in (a) occur along this line due to switching between $A$ and $B$. **e**, Coercivity of the hysteretic switching calculated within the Master equation approach (see supplementary information). The theoretical trend agrees well with the experimental switching dynamics in (c).

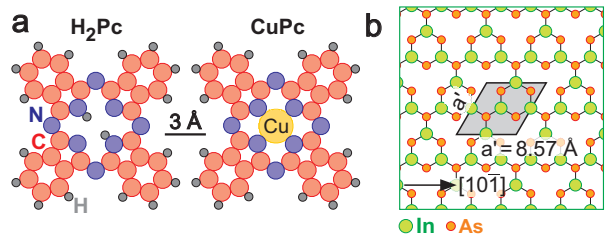

**a** H₂Pc        CuPc

N
C       3 Å       Cu
H

**b**

a' = 8.57 Å

[101̄]

In ⬤  As ⬤

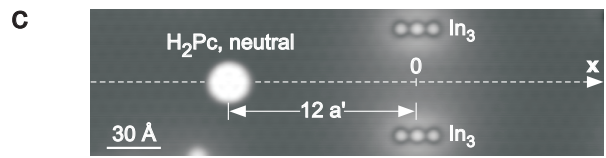

**c** H₂Pc, neutral        In₃

0                    x

30 Å    ⟵ 12 a' ⟶

In₃

**d**

a'  [101̄]

30 Å        H₂Pc charged −1

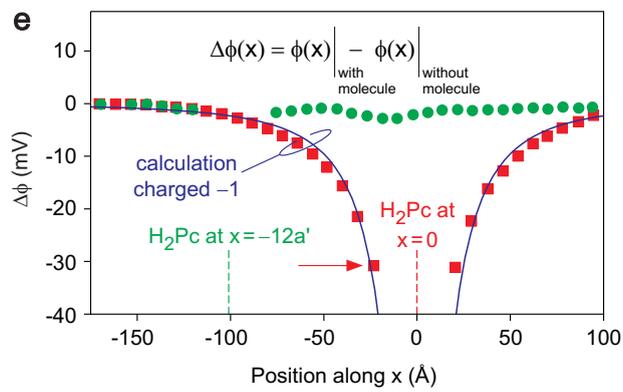

**e**

$$\Delta\phi(x) = \phi(x)\bigg|_{\text{with molecule}} - \phi(x)\bigg|_{\text{without molecule}}$$

calculation charged −1

H₂Pc at x = −12a'        H₂Pc at x = 0

Δφ (mV) axis: 10, 0, −10, −20, −30, −40

Position along x (Å): −150, −100, −50, 0, 50, 100

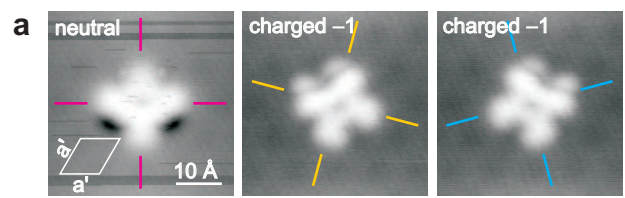

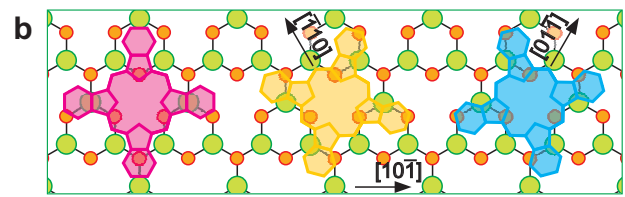

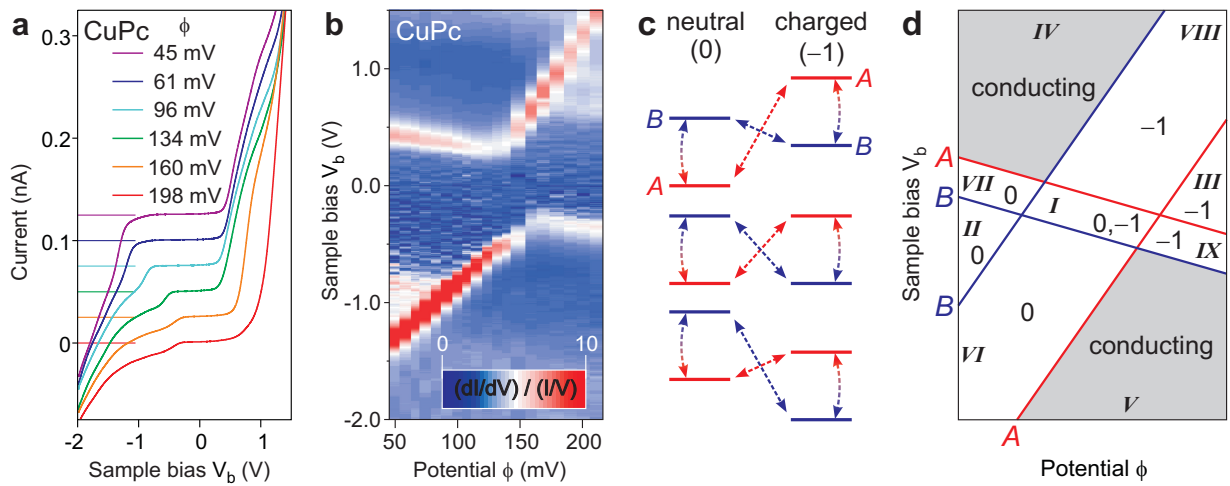

**a**

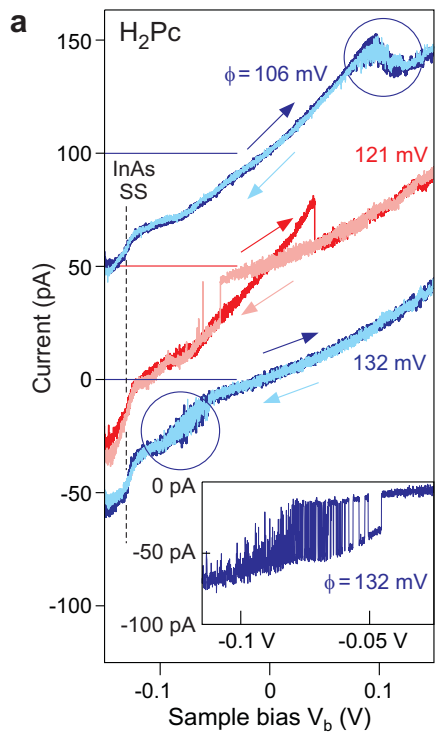

H₂Pc

$\phi$ = 106 mV

121 mV

132 mV

InAs
SS

0 pA
-50 pA
-100 pA

$\phi$ = 132 mV

-0.1 V    -0.05 V

Current (pA)

150

100

50

0

-50

-100

Sample bias V_b (V)

-0.1    0    0.1

**b**

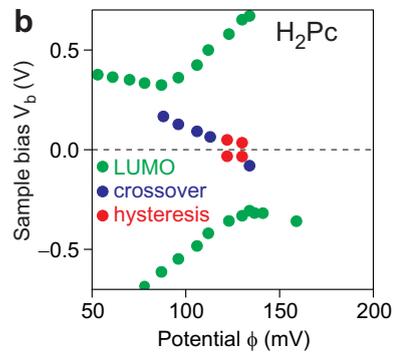

H₂Pc

Sample bias V_b (V)

0.5

0.0

-0.5

LUMO
crossover
hysteresis

Potential $\phi$ (mV)

50    100    150    200

**c**

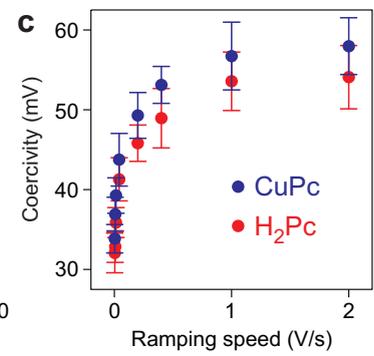

Coercivity (mV)

60

50

40

30

CuPc
H₂Pc

Ramping speed (V/s)

0    1    2

**d**

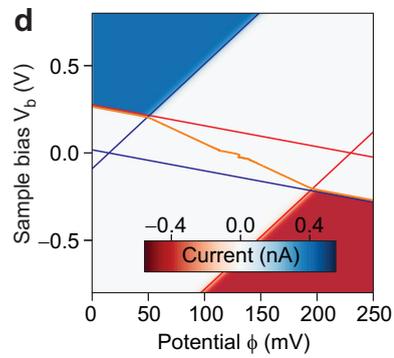

Sample bias V_b (V)

0.5

0.0

-0.5

Current (nA)

-0.4    0.0    0.4

Potential $\phi$ (mV)

0    50    100    150    200    250

**e**

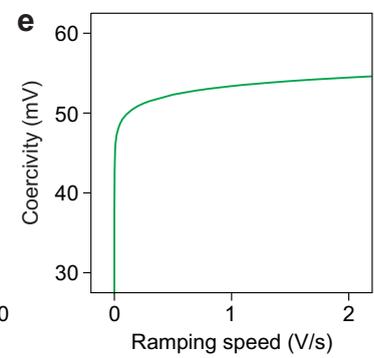

Coercivity (mV)

50

40

30

Ramping speed (V/s)

0    1    2